\documentclass[11pt]{article}
\usepackage{fa2025}
\usepackage{amsmath}
\usepackage{cite}
\usepackage{url}
\usepackage{graphicx}
\usepackage{color}
\usepackage{siunitx}
\usepackage[utf8]{inputenc}
\usepackage{svg}

\title{Automated Data Curation for Self-Supervised Learning in Underwater Acoustic Analysis}

\multauthor
{Hilde I Hummel$^{1*}$ \hspace{1cm} Sandjai Bhulai$^2$ \hspace{1cm} Burooj Ghani$^3$} { \bfseries{Rob van der Mei$^1$ }\\
  $^1$ Center of Mathematics and Computer Science, Amsterdam, Netherlands\\
$^2$ Vrije Universiteit, Amsterdam, Netherlands\\
$^3$  Naturalis Biodiversity Center, Leiden, Netherlands
\correspondingauthor{h.i.hummel@cwi.nl}{Hummel et al.}
}

\begin{document}

\maketitle
\begin{abstract}
The sustainability of the ocean ecosystem is threatened
by increased levels of sound pollution, making monitoring crucial to
understand its variability and impact. \textit{Passive acoustic monitoring} (PAM) systems collect a large amount of underwater sound recordings, but the large volume of data makes manual analysis impossible, creating the need for automation. Although machine learning offers a potential solution, most underwater acoustic recordings are unlabeled. Self-supervised learning models have demonstrated success in learning from large-scale unlabeled data in various domains like computer vision, Natural Language Processing, and audio. However, these models require large, diverse, and balanced datasets for training in order to generalize well. To address this, a fully automated self-supervised data curation pipeline is proposed to create a diverse and balanced dataset from raw PAM data. It integrates \textit{Automatic Identification System} (AIS) data with recordings from various hydrophones in the U.S. waters. Using hierarchical k-means clustering, the raw audio data is sampled and then combined with AIS samples to create a balanced and diverse dataset. The resulting curated dataset enables the development of self-supervised learning models, facilitating various tasks such as monitoring marine mammals and assessing sound pollution. 
\end{abstract}

\section{Introduction}\label{sec:introduction}
The increasing levels of sound pollution threaten the preservation of ocean ecosystems, necessitating the monitoring of underwater sounds \cite{hummel2024survey}. \textit{Passive Acoustic Monitoring} (PAM) systems are globally deployed and collect a vast amount of diverse underwater sound recordings. The complexity of the marine environment, combined with the large volume of data, makes manual analysis impractical. As a result, annotating such data is time-consuming and expensive, leaving most PAM recordings unlabeled \cite{hummel2024survey}. Although \textit{machine learning} (ML) could potentially facilitate automatic analysis, the performance of supervised methods is limited by the scarce label availability. \textit{Self-Supervised Learning} (SSL) has successfully handled large-scale unlabeled data in domains like computer vision, Natural Language Processing \cite{gui2024survey}, and audio \cite{liu2022audio}. However, the performance of SSL models is reduced when they are optimized on uncurated data \cite{jose2024dinov2}. This makes balancing of the training data by data curation a key aspect for SSL models to generalize on downstream tasks. Despite the widespread availability of uncurated PAM data, no prior research has focused on curating these data to support the development of SSL-driven models for automatic underwater acoustic analysis. To address this gap, this article presents a fully automated pipeline for the curation of large-scale, freely accessible PAM recordings. This pipeline integrates \textit{Automatic Identification System} (AIS) data with PAM data to balance both the ship distribution and the raw audio distributions. The key contributions of this paper are: 
\begin{itemize}
    \item A simple AIS curation method based on ship occurrence; 
    \item An online hierarchical clustering approach for raw large-scale PAM data curation.
\end{itemize}

The curation method is evaluated by training an SSL algorithm on the curated dataset. The embedding space will be evaluated on a downstream task. The learned embeddings serve as input for the linear regression-based classification, to classify ship types from the reference datasets Deepship \cite{irfan2021deepship} and ShipsEar \cite{santos2016shipsear} containing underwater recordings of ships labeled by the type of ship.

\section{Related Work}\label{sec:relatedwork}
\subsection{SSL in underwater acoustics}
Several previous studies have shown the potential of SSL in the automatic recognition of ship types \cite{hummel2024survey}. With no large, curated underwater acoustic dataset publicly available yet, \cite{xu2023self} and \cite{xu2023selfmix} proposed pretraining on AudioSet applying a mix-up strategy. Next, a Swin Transformer encoder is optimized with masked log-Melspectrogram, while a decoder is optimized to reconstruct the full spectrogram. A high masking rate was needed because of the limited contrast in these spectrograms. The fine-tuning on Deepship \cite{irfan2021deepship} yielded an accuracy of 80\% - 86\%. However, because of the complexity of the marine environment, tasks related to acoustics above the water may not be representative of the marine environment. To address this, \cite{feng2024masking} proposed a generative transformer model pre-trained solely on underwater acoustic data. To reduce the GPU demand during training, they proposed hierarchical token-convolutions and masked these for reconstruction. Deepship model pre-training and fine-tuning on both Deepship and ShipsEar \cite{santos2016shipsear} improved performance over a supervised transformer baseline. However, their method is limited in real-world applications because of the lack of diversity of ocean environments in their training data. Our study overcomes this issue by representing a data curation pipeline to extract a diverse and balanced unlabeled dataset for SSL algorithms.

\subsection{Data curation methods}
SSL models need a balanced and diverse dataset to compete with supervised methods. The simple scraping of raw data from the web results in an unbalanced dataset, creating the need for data curation \cite{jose2024dinov2}. Previous research proposed an automatic data curation pipeline for data curation from images scraped from the web \cite{oquab2023dinov2}. They suggested converting both a labeled curated dataset and an unlabeled uncurated dataset to the embedding space and performed deduplication by removing near-duplicate images in the uncurated dataset. Finally, they retrieved images from the uncurated dataset by calculating the relative distance to the curated samples. Although they state that this method is self-supervised, it still requires a large labeled dataset for matching. In a follow-up article, \cite{vo2024automatic}, they proposed a hierarchical KMeans algorithm to retrieve images from an uncurated dataset. Here, they overcome the problem that web-scraped big data is mostly long-tailed. Sampling from traditional KMeans does not encounter this, and therefore, hierarchical KMeans clustering and sampling are proposed. In addition, the DINOv2 algorithm was applied to text-image pairs \cite{jose2024dinov2}. In this paper, they performed data curation. Here, they aligned the images to the captions and sampled from the aligned captions to create the image dataset. The addition of metadata for data curation has also been proposed for audio-visual video representations in \cite{lee2021acav100m}. In this work, they tried to maximize the estimated mutual information between the video imagery and the corresponding audio. Both studies showed that the incorporation of metadata for data curation can improve the quality of the curated data. 


\section{Methods}\label{sec:methods}
\subsection{PAM data}
\begin{figure}[ht]
 \centerline{
 \includegraphics[width=7.8cm]{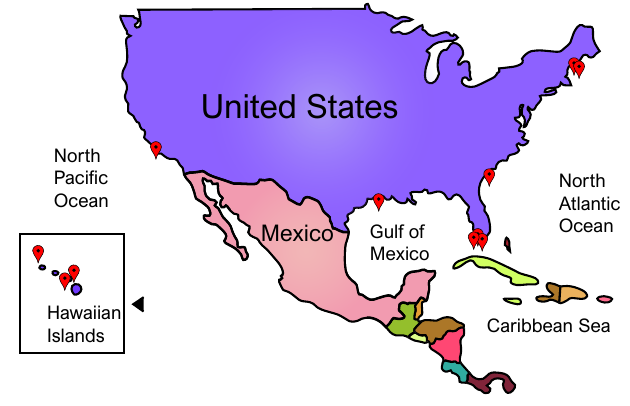}}
 \caption{The locations of the hydrophones from the web scraped PAM data, indicated in red pointers.}
 \label{fig:hydrophones}
\end{figure}

The raw audio data is collected from NOAA\footnote{\url{https://console.cloud.google.com/storage/browser/noaa-passive-bioacoustic;tab=objects?inv=1&invt=AbnmXQ&prefix=&forceOnObjectsSortingFiltering=false}}, selecting hydrophones which started recording in 2023 or later. All these PAM audio recordings are combined into the resulting dataset $\mathcal{D}$. In total, this combination covers the duration of 8 years, 6 months, 9 days, 15 hours, 19 minutes, and 49 seconds from 11 individual hydrophones. These hydrophones are visualized in Figure \ref{fig:hydrophones}.

\subsection{AIS data}
\begin{figure}[ht]
 \centerline{
 \includegraphics[width=7.8cm]{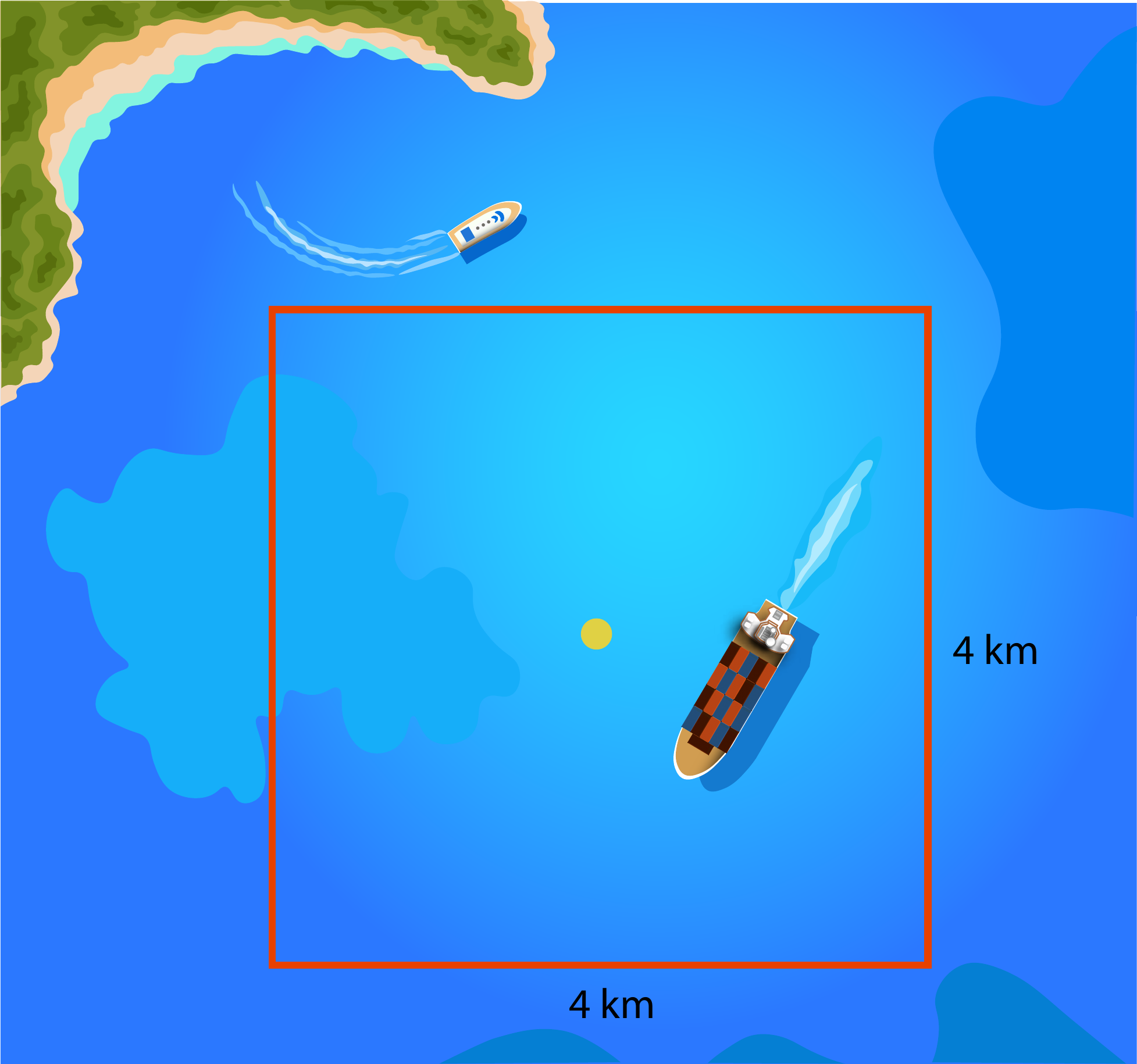}}
 \caption{The defined range of the selected hydrophones to align raw PAM recordings with the AIS pulses.}
 \label{fig:range}
\end{figure}

Track records of ships are recovered by AIS. This system captures information about ship characteristics, the location, and movements of almost all ships worldwide. The NOAA also provides freely accessible AIS data\footnote{\url{https://marinecadastre.gov/accessais/}} covering the same regions as raw PAM data. The raw PAM data were aligned with the AIS data by drawing a 4 km $\times$ 4 km square with the hydrophone in the center (Figure \ref{fig:range}). This square defines the range of the hydrophone, and therefore, every AIS pulse within this range was considered recorded. The data set containing the AIS pulses will be referred to as $\mathcal{A}$. All recordings, with AIS information, were gathered to create the dataset $\mathcal{D}_s$. 

\subsection{AIS data curation}
The low contrast of radiated ship noise against background noise makes it difficult for simple machine learning methods to encounter ship noise in a large volume of data. To ensure a diverse representation of the acoustic profiles of the ships in the curated dataset, the AIS data $\mathcal{A}$ is also curated. A simple approach inspired by \cite{jose2024dinov2} is proposed. Here, they proposed to curate the aligned caption to curate an image dataset from a caption-image pair dataset. In this work, the AIS data are aligned with the PAM data, creating a dataset $\mathcal{D}_s$ holding the aligned PAM recordings and a dataset $\mathcal{A}$ holding the aligned AIS pulses. These pulses are then sampled based on the occurrence of individual ships, using a threshold value $t$. This threshold is defined based on the incidents of individual ships in the aligned PAM dataset $\mathcal{D}_s$. Ships with a lower incidence than $t$ will not be sampled and kept complete in the curated dataset, while ships that exceed $t$ are sampled according to their sample probability. This probability is inversely proportional to the occurrence of the ship in $\mathcal{D}_s$, the higher the occurrence, the lower the sampling probability and vice versa. Here, the objective is to balance the long-tailed distribution of $\mathcal{D}_s$ given $\mathcal{A}$, resulting in a more balanced dataset $\mathcal{D}_s^*$:

\begin{equation}
    \mathcal{D}_s^* \leftarrow f(\mathcal{D}_s;\mathcal{A},t)
\end{equation}

\subsection{Audio preprocessing}
First, the raw audio was resampled to a sample rate of 16 kHz to ensure a common sample rate for all recordings. Next, the audio was windowed using a 10-second window size without overlap. From this windowed audio, embeddings were generated using the model presented in \cite{hummel2025computation}. In this research, they state that this model is optimized on a large quantity of unlabeled underwater acoustic data and generates generalized embeddings with dimension 2048. 

\subsection{PAM data curation}
The objective of the PAM data curation method is to retrieve a dataset with the target distribution $U$. This distribution is defined as a uniform distribution with the support of the distribution $P$, where $P$ represents the distribution of dataset $\mathcal{D}$. The data curation method is inspired by the method proposed by \cite{vo2024automatic}. The entire dataset $\mathcal{D}$ is employed to optimize a hierarchical KMeans model using the resampling-clustering technique. This model consists of four levels with cluster sizes of $[6000,400,40,10]$. Here, the lower cluster levels focus on detailed information, while the upper levels capture global features. Audio samples of 10 seconds are then drawn from each hierarchical layer. Given the large size of the dataset $\mathcal{D}$, the hierarchical KMeans is optimized in a streaming manner, which approximates the optimal solution \cite{sculley2010web}. After training, the optimal HKmeans is utilized to sample the data from each cluster level. A maximum of target size $N$ samples is selected from the first datastream based on the relative distance of the datapoints to the assigned cluster centers.  As additional datastreams are processed, once the target number of $N$ samples is exceeded, the samples with the greatest distance to the cluster center are replaced by those closer to the center.  




\subsection{SSL algorithm training and evaluation}
The proposed method is evaluated by training the Data2vec base framework\cite{baevski2022data2vec}. The base model pretrained on speech is completely fine-tuned using the curated dataset. This dataset is defined as:
\begin{equation}
    \mathcal{D}^* = \mathcal{D}_a + \mathcal{D}_s^*
\end{equation}
The model is optimized using a batch size of 64, where each input sample for the student model is masked for 15\% while the teacher model receives the full audio. The teacher model weights $\Delta$ keep track of the student model weights $\theta$ by:
\begin{equation}
    \Delta = \tau \Delta + (1-\tau)\theta
\end{equation}
The parameter $\tau~( \tau_0 \leq \tau \leq \tau_e)$ is gradually increased from $\tau_0=0.999$ to $\tau_e=0.9999$ over 20 updates. The performance of the model is compared by training the same framework on a randomly curated dataset of the same size. A simple logistic regression is optimized using the learned embeddings for the classification of the type of ships.



\section{Results}\label{sec:results}
\subsection{Data curation}
\begin{figure}[ht]
 \centerline{
 \includegraphics[width=7.8cm]{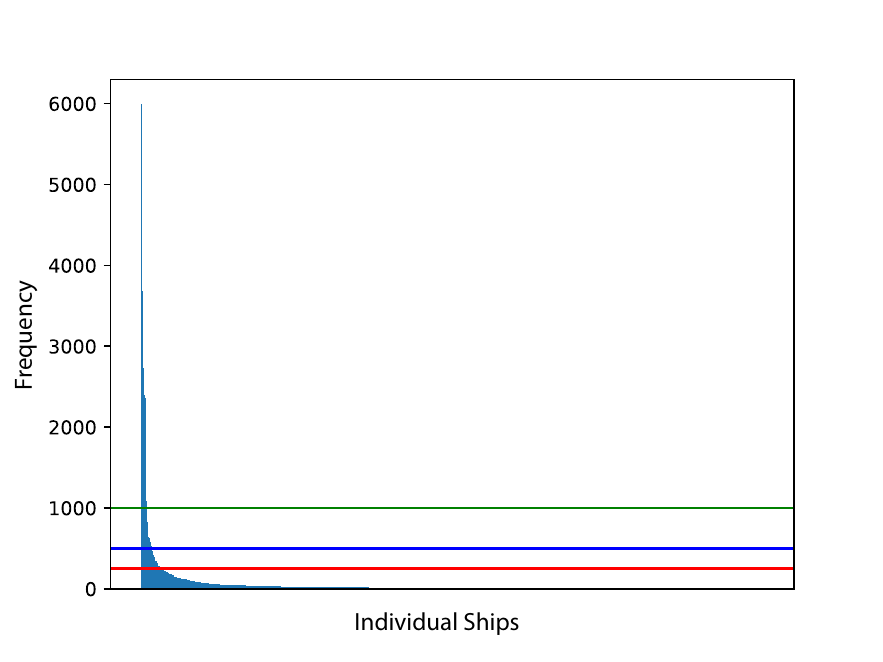}}
 \caption{ AIS distribution of individual ships in datset $\mathcal{D}_s$ with threshold value $t=250$ in red, $t=500$ in blue, $t=1000$ in green}
 \label{fig:AISdistribution}
\end{figure}

For the AIS curation method, the optimal threshold value $t$ is defined to correspond to the knee of the skewed distribution of $\mathcal{D}_s$ \cite{jose2024dinov2}. Figure \ref{fig:AISdistribution} illustrates the number of 10-second audio windows per individual ship, revealing a skewed distribution. The optimal threshold, aligning with the knee of the distribution, is around 250. For this reason, the threshold value $t$ was set to 250, resulting in a dataset $\mathcal{D}_s^*$ capturing 25,021 audio samples. 

In addition to AIS samples, 323,532 samples were selected during PAM data curation. These samples were combined to create the dataset $\mathcal{D}^*$ holding roughly 970 hours of PAM recordings.





\subsection{SSL results curated dataset}
The proposed method is evaluated using the labeled benchmark datasets Deepship \cite{irfan2021deepship} and ShipsEar \cite{santos2016shipsear}. Both datasets hold ship types recorded in two distinct regions. Table \ref{tab:Accuracy} presents the accuracy scores of the Data2Vec model trained on curated and random datasets. The results indicate that the curated model outperforms the random model on both benchmark datasets. However, the performance gain is smaller for ShipsEar, likely due to environmental factors. The ocean environment of ShipsEar is quite different from the environment of the hydrophones in the raw PAM data (Figure \ref{fig:hydrophones}), making classification more challenging. In contrast, the ocean environment of Deepship is more comparable with the PAM data, leading to a performance increase of more than 7\%. 

\begin{table}[!h]
 \caption{Accuracy scores ship type classification}
 \begin{center}
 \begin{tabular}{|l|l|l|}
  \hline
   & ShipsEar & Deepship \\
  \hline
  Random  & 51.98\% & 49.16\%\\
  Curated & 53.11\%& 56.72\%\\
  \hline
 \end{tabular}
\end{center}
 \label{tab:Accuracy}
\end{table}






\section{Conclusion}\label{sec:conclusion}
This work describes the first automatic data curation pipeline to curate large web-scraped PAM data. The study demonstrates that curation is a key aspect in extracting accurate SSL model representations from unlabeled underwater recordings. Although this work focuses on data curation, more research is still required on SSL methods applied to underwater acoustics. Due to the stationarity of the data, the masking strategy may be suboptimal, and the results may benefit from a more contrastive approach. In addition to the raw web-scraped PAM data used in this study, more diverse data is publicly available. Expanding the diversity by incorporating more hydrophones from various regions would make the representations more robust for various other underwater acoustic-related tasks. This offers the possibility of training large-scale SSL models from scratch, enabling more advanced underwater acoustic applications.

\bibliography{fa2025_template}

%
%
%

\end{document}